\documentclass[referee]{cjaa}           % referee version: for submission
\usepackage{graphicx}                   %for PS/EPS graphics inclusion, new
\input{epsf.sty}                        %for PS/EPS graphics inclusion, old
\input{psfig.sty}                       %for PS/EPS graphics inclusion, old

\begin{document}

   \title{Chaotic Feature in the Light Curve of 3C 273}

   \volnopage{Vol.0 (200x) No.0, 000--000}      %%preserved for Editor. DOn't remove!
   \setcounter{page}{1}          %%starting page, preserved for Editor. DOn't remove!

   \author{Lei Liu \inst{}\mailto{}}

   \offprints{Lei Liu}                   %% is disabled in fact

   \institute{Institute of Particle Physics, Huazhong Normal University, Wuhan 430079,
              China\\ \email{liuphy@mail.ccnu.edu.cn}}

   \date{Received~~2006 month day; accepted~~2006~~month day}

   \abstract{
Some nonlinear dynamical techniques, including state-space
reconstruction and correlation integral, are used to analyze the
light curve of 3C 273. The result is compared with a chaotic model.
The similarities between them suggest that there is a
low-dimensional chaotic attractor in the light curve of 3C 273.
   }

%   \authorrunning{Lei Liu }            %author_head in even pages
%   \titlerunning{Chaotic Feature in the Light Curve of 3C 273}  % title_head in odd pages

   \maketitle

\section{Introduction}           %% first-level sections will be auto-capitalized
%\hspace{15pt}%                   %% preserved for Editor
Since its discovery by Smith and Hoffleit (1963), the light curve of
3C 273 plays an important role to understand the nature of the
quasar. Although it has been subjected to extensive analysis, there
is no accepted method of extracting the information from the light
curve. The divergence in result of analyzing the light curve ranges
from multi-periodic behavior (Kunkel 1967; Jurkevich 1971; Sillanpaa
et al.~1988; Lin 2001) to a purely random process (Manwell \& Simon
1966, 1968; Terrell \& Olsen 1970; Fahlman \& Ulrych 1975). Whatever
does this $seemingly$ random light curve tell us? Could this seeming
randomness be another kind of behavior other than the multi-periodic
behavior or a purely random process?

  In 1963, Edward Lorenz published his monumental work entitled
$Deterministic~Nonperiodic~Flow$. In this paper, he found a strange
behavior which can appear in the deterministic nonlinear dissipative
system. This behavior seems random and unpredictable and is called
$Chaos$. Chaotic behavior is not multi-periodic because it has a
continuous spectrum. Useful information can not always be extracted
from power spectrum of chaotic signal. Chaotic behavior is also not
random because it can appear in a completely deterministic system.
The concept of $attractor$ is often used to describe the chaotic
behavior. As the dissipative system evolves in time, the trajectory
in state space will head for some final region. We call this final
region attractor. The attractor may be an ordinary Euclidean object
or a fractal (Feder 1988) which has non-integer dimension and often
appears in the state space of chaotic system. For many practical
systems, we may not know in advance the required degrees of freedom
and of course can not measure all the dynamic variables. How can we
discern the nature of the attractor from the available experimental
data? In 1980, Packard et al. introduced a technique which can be
used to reconstruct state-space attractor from the time series data
of a $single$ dynamical variable. A practical algorithm subsequently
introduced by Grassberger and Procaccia (1983) can be used to
determine the dimension of the attractor embedded in the new state
space. These techniques constitute a useful diagnostic method of
chaos in the practical system.

  3C 273 may be a complex nonlinear dissipative system. If so, the
complex light variation of 3C 273 may be chaotic. Here we use the
techniques introduced by Packard et al. and Grassberger et al. to
investigate whether there is a chaotic attractor in the light curve
of 3C 273. The paper is arranged as follows: in Sect.~2, a brief
introduction to the method for diagnosing chaos; in Sect.~3, we
apply this method to a chaotic model and 3C 273, and then the
results are compared; in Sect.~4, some discussions are given.

\section{Method of analysis}
%\hspace{15pt}%                   %% preserved for Editor
  The state-space reconstruction technique (Packard et al.~1980) which
is based on the notion that the attractor of a multi-dimensional
dissipative system can be often reconstrcted by using the time
series data of a $single$ variable. Since a detailed presentation of
this technique is available in many places ( see , for example,
Hilbron 1994; Abarbanel et al.~1993; Sprott 2003), we just give a
brief introduction below. Let $X_{1},X_{2},\ldots,X_{N}$ be
measurements of a physical variable at the times
$t_i=t_0+(i-1)\Delta{}t,i=1,\ldots,N.$ From this sequence one can
construct a set of $d$-dimensional vectors $\mathbf{v_i},
i=1,\ldots,N-(d-1)T$ of the form
\begin{equation}
\mathbf{v_i}=(X_i,X_{i+T},X_{i+2T},\ldots,X_{i+(d-1)T}) ,
\end{equation}
where $T$, called the $time~delay$, is an integral multiple of
$\Delta{}t.$ We assume that the real attractor in the full state
space of the system can be reconstructed from the time-delayed
vectors $\mathbf{v_i}$ moving in the $d$-dimensional state space.
$d$ is often called $embedding~dimension$. This assumption works
well when embedding dimension becomes greater than about twice the
dimension of the real attractor (Sprott 2003).

  The correlation integral (Grassberger \& Procaccia 1983; see also
Hilbron 1994; Abarbanel et al.~1993; Sprott 2003) can be used to
determine the attractor dimension. We define this to be
\begin{equation}
C(r)=\frac{1}{(N-k)(N-k-1)}\sum_{i=1}^{N-k}\sum_{j=1,j \neq
i}^{N-k}\theta(r-|\mathbf{v_i}-\mathbf{v_j}|)
\end{equation}
where $k=(d-1)T$, and $\theta(x)$ is the Heaviside function,
\begin{equation}
 \theta(x)=  \left\lbrace \begin{array}{ll} 1 ~~~~~~x \geq 0\\
                                         0 ~~~~~~x<0
                       \end{array} \right.
\end{equation}
For the dissipative system , $C(r)$ behaves as a power of $r$ for
small $r$,
\begin{equation}
C(r)\propto r^{D}
\end{equation}
where $D$ is called $correlation~dimension$. Strictly speaking , $D$
is not attractor dimension, but very close to it (Grassberger and
Procaccia 1983). Thus attractor dimension can be estimated by using
the correlation dimension. Note that for large values of $r$ the
finite size of the attractor makes $C(r)$ ``saturate'' at 1 and for
small values of $r$ the finite number of data points causes $C(r)$
to be close to zero. Thus, the curve of $log_{10}~C(r)$ versus
$log_{10}~r$ is approximately a straight line just in the
intermediate region , as in Fig.~1.

  What we first do in practice is to compute the correlation dimension $D$
by using Eq.~(2) and Eq.~(4). As the value of $D$ depends on the
delay time $T$ and the embedding dimension $d$, we then plot $D$
versus $d$ for different values of $T$, as in Fig.~2. If there is a
chaotic attractor, $D$ should be independent of $d$ until d becomes
greater than some value defined by $d_{sat}$. For some special value
of T, $d_{sat}$ is about twice the value of saturation correlation
dimension $D_{sat}$, that is the value independent of $d$ (Sprott
2003). Thus the saturation dimension $D_{sat}$ which corresponds
with this special value of $T$ is the attractor dimension which we
expect to find by using a single time series data.

\section{Data analysis}
%\hspace{15pt}%                   %% preserved for Editor
  First, we apply the method described in Sect.~2 to Lorenz model
which was introduced by E. Lorenz (1963) to model convection in the
atmosphere. It treats the fluid system (say, the atmosphere)as a
fluid layer that is heated at the bottom (due to the sun's heating
the earth's surface, for example) and cooled at the top. A detailed
derivation of the equations of the Lorenz model can also be found in
many textbooks (see, for example, Hilbron 1994). Here we just give
the result,
\begin{eqnarray*}
dx/dt&=&\sigma(y-x) \\
dy/dt&=&-xz+\gamma x-y \\
dz/dt&=&xy-bz.
\end{eqnarray*}
It is important to stress that the Lorenz model introduced here is
not treated as a model of 3C 273. We just use it to produce a set of
chaotic time series and show what would happen when the method of
chaos diagnostication is used to analyze a set of chaotic time
series. We choose the parameters $\sigma=10, \gamma=28$ and $b=8/3$
which are used by many authors to produce a set of chaotic solution
of Lorenz equations. We use four-order Runge-Kutta method to solve
the equations with the initial conditions $x_0=0,y_0=-0.01,z_0=9,$
and generate 292 data points representing $x(t)$ at equally spaced
time intervals of $\Delta{}t=0.05.$ Note that the 292 data points
are generated from t=50 to assure the trajectory is on the
attractor.

\begin{figure}
    \centering
   \includegraphics[width=120mm,height=60mm]{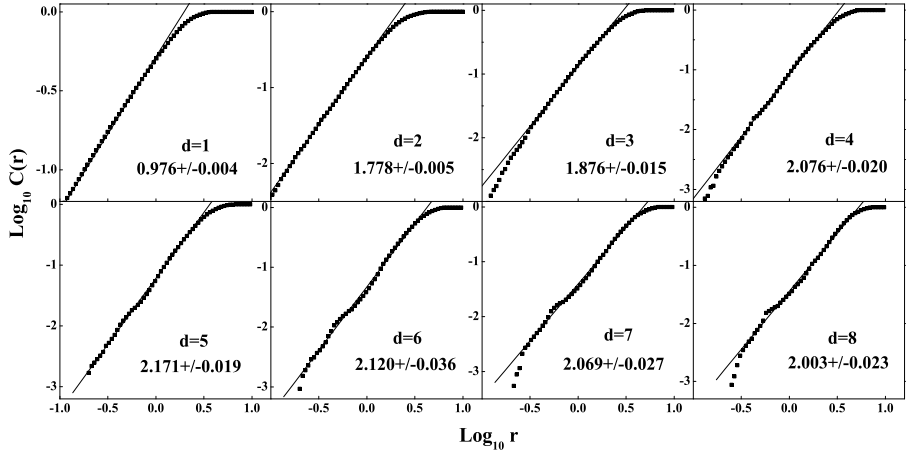}
   \caption{Correlation integral $C(r)$ of Lorenz model for time delay $T=20$ and the embedding
dimension $d=1,2,\ldots,8$ on doubly logarithmic scales. In each
panel, the scale of C(r) is arbitrary and the scale of r is same. We
also give the slope fitted by using Eq.~(4) in each panel.}
   \label{Fig:plot1}
\end{figure}

\begin{figure}
    \centering
   \includegraphics[width=120mm,height=80mm]{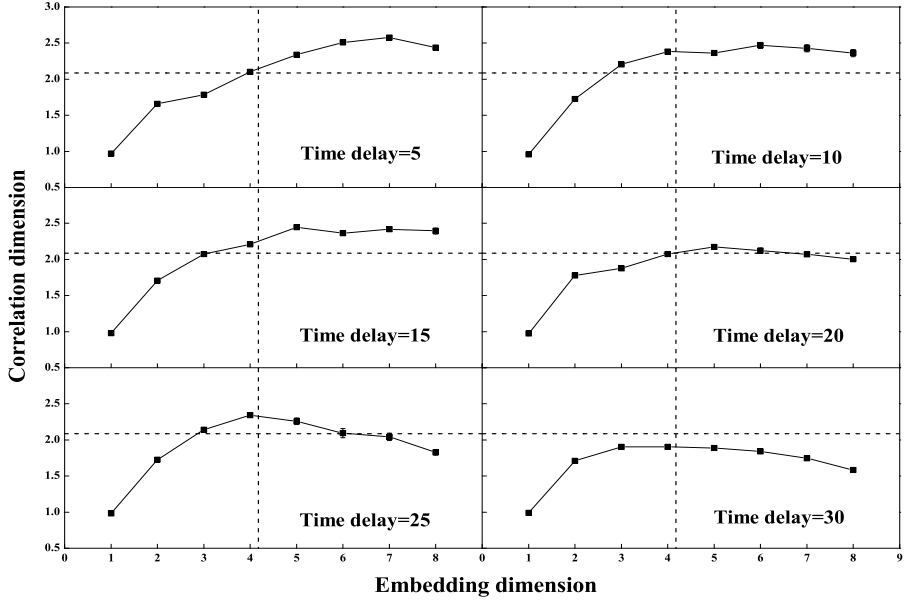}
   \caption{Correlation dimension $D$ versus the embedding dimension $d$ for Lorenz model.
   Each panel has the same scale of $D$ and also the same scale of $d$.}
   \label{Fig:plot1}
\end{figure}
  The 292 data points are analyzed by using Eq.~(2) and Eq.~(4). The
results of $log_{10}~C(r)$ plotted against $log_{10}~r$ for time
delay $T=20$ are shown in Fig.~1. We can see that in the
intermediate region there is a approximately straight line. We also
plot $log_{10}~C(r)$ versus $log_{10}~r$ for some other $T$ and find
that the curves are also approximately straight lines in the
intermediate region as in Fig.~1. The results of correlation
dimension $D$ versus embedding dimension $d$ for different time
delay $T$ are shown in Fig.~2. Now we carefully compare these curves
for different $T$, and find out that the value of $T$ which makes
$d_{sat} \approx 2D_{sat}$ to estimate the attractor dimension. Note
that when $T=25$ and $30$, we can't find a good plateau, where $D$
is independent of $d$, to estimate the attractor dimension. When
$T=5$, the $d_{sat}$ is close to $6$. And $D_{sat} \approx 2.508 \pm
0.026$ obtained by averaging the values of correlation dimension for
$d=6,7,8$. When $T=10,15$ and $20$, the $d_{sat}$ is close to $4$.
And the corresponding $D_{sat} \approx 2.401 \pm 0.039,2.367 \pm
0.031$ and $2.088 \pm 0.025$ obtained by averaging $D$ for
$d=4,5,6,7,8$. Thus the best value which makes $d_{sat} \approx
2D_{sat}$ is 20. The value of $D_{sat}$ estimated from the curve of
$T=20$ is about $2.088 \pm 0.025$. Thus, the attractor dimension is
about $2.088 \pm 0.025$. Two crossing dash lines representing
$D_{sat}=2.088$ and $d_{sat}=4.176$ are put in Fig.~2. From the two
lines, we can not only easily see why $T=20$ is the special value
which we choose to estimate the attractor dimension but also find
that there is a downward trend in the value of $D_{sat}$ as $T$
increases. By using the full state space vectors and a large amount
of data, one can find a more accurate value of the correlation
dimension of Lorenz chaotic attractor which equals $2.068\pm 0.086$
(Sprott 2003). Our result is very close to it.

  From the analysis of Lorenz model, we know that the curve of
$log_{10} C(r)$ versus $log_{10} r$ is approximately a straight line
in the intermediate region for the dissipative system with an
attractor. Thus the correlation dimension can be estimated from the
slope of the straight line. Moreover, for the dissipative system,
the correlation dimension $D$ will be independent of embedding
dimension $d$ as $d$ increases and some special value of delay time
$T$ which makes $d_{sat} \approx 2D_{sat}$ can be found out.
\begin{figure}
   \centering
   \includegraphics[width=120mm,height=60mm]{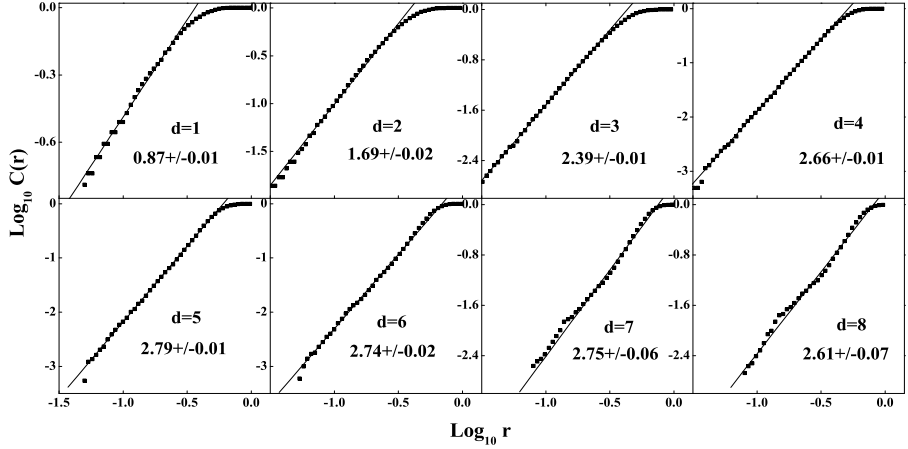}
   \caption{Correlation integral $C(r)$ of 3C 273 for time delay $T=30$ and the embedding
dimension $d=1,2,\ldots,8$ on doubly logarithmic scales. In each
panel, the scale of C(r) is arbitrary and the scale of r is same. We
also give the slope fitted by using Eq.~(4) in each panel.}
   \label{Fig:plot1}
\end{figure}

\begin{figure}
   \centering
   \includegraphics[width=120mm,height=80mm]{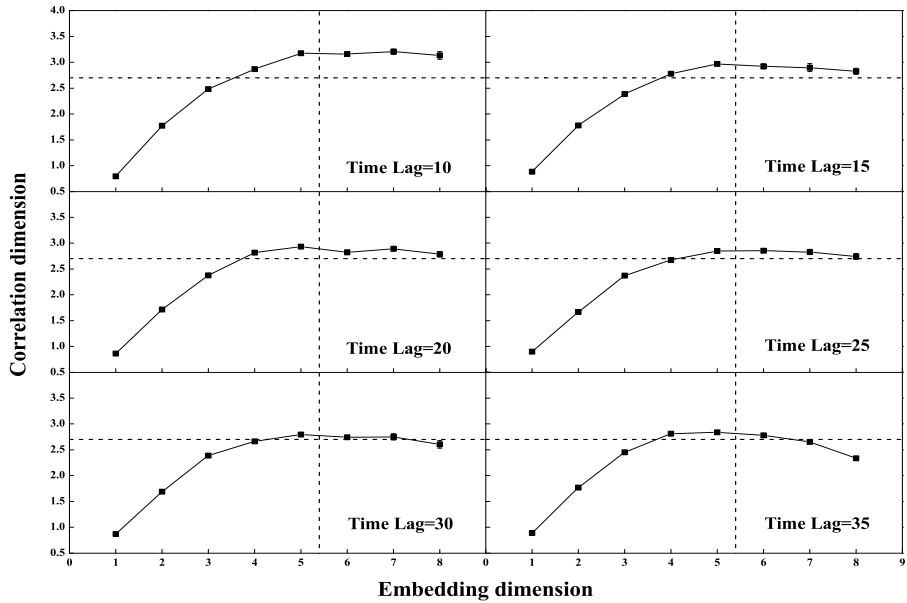}
   \caption{Correlation dimension $D$ versus the embedding dimension $d$ of 3C 273.
    Each panel has the same scale of $D$ and also the same scale of $d$.}
   \label{Fig:plot1}
\end{figure}

  Now we analyze the 292 data points of the light curve of 3C 273 in
the form of 100-day means (Kunkel 1967). The results of
$log_{10}~C(r)$ plotted against $log_{10}~r$ for time delay $T=30$
are shown in Fig.~3. The curves are also approximately straight
lines in the intermediate region as in Fig.~1. The results of
correlation dimension $D$ plotted against $d$ for different time
delay $T$ are shown in Fig.~4. As in Fig.~2, when $T=35$, it is
difficult to find a good plateau to estimate the attractor
dimension. When $T=10,15,20,25$ and $30$, the $d_{sat}$ is close to
$5$. And the corresponding $D_{sat} \approx 3.17 \pm 0.05,2.90 \pm
0.05,2.86 \pm 0.04,2.82 \pm 0.04$ and $2.72 \pm 0.04$ obtained by
averaging $D$ for $d=5,6,7$ and $8$. Thus the best value which makes
$d_{sat} \approx 2D_{sat}$ is 30 and the attractor dimension is
estimated to be about $2.72 \pm 0.04$. Two crossing dash lines in
Fig.~4 represent $D_{sat}=2.7$ and $d_{sat}=5.4$. By using the two
lines, we also easily find that why $T=30$ is special and that there
is a downward trend in the value of $D_{sat}$ as $T$ increases.

  At last, we compare Lorenz model with 3C 273. First, the curve of
$log_{10} C(r)$ versus $log_{10} r$ is approximately a straight line
in the intermediate region both in the Lorenz model and 3C 273.
Second, the plateau, where the embedding dimension is independent of
the correlation dimension, can be found at some values of time delay
T both in the Lorenz model and 3C 273. Third, the special time delay
$T$ which makes $d_{sat} \approx 2D_{sat}$ can be found out both in
the Lorenz model and in the 3C 273. Fourth, there is a downward
trend in the value of $D_{sat}$ as $T$ increases both in the Lorenz
model and in the 3C 273. The similarities between them strongly
suggest that there is a low-dimensional chaotic attractor in the
light curve of 3C 273.

\section{Discussion}
%\hspace{15pt}%                   %% preserved for Editor
The nature of AGNs is still an open question. The study of the
variation in the light curve of AGNs is expected to yield valuable
information about the nature of AGNs. 3C 373 is known as the
brightest AGN. In this paper, the state-space reconstruction and
correlation integral are used to analyze the light curve of 3C 273.
The result is compared with Lorenz model. The similarity between
them strongly suggests that there is a low-dimensional chaotic
attractor in the light curve of 3C 273. Thus the variation of the
light curve of 3C 273 has the nonlinear dynamical origin. It can not
be interpreted as multi-periodic behavior or a purely random
process. The evidence of chaotic behavior which we show tells that
the concepts of nonlinearity may be helpful to understand the nature
of the variation in the light curve of AGNs.

It is interesting to compare 3C 273 with other sources. Misra et al.
(2004, 2006) have analyzed the X-ray light curves of GRS 1915+105 by
using the same method along with surrogate data analysis and find
the evidence which is provided for a non-linear limit cycle origin
of one of the low frequency QPO detected in the source, while some
other types of variability could be due to an underlying low
dimensional chaotic system. The chaotic behavior found in the
microquasar GRS 1915+105 and the quasar 3C 273 implies that the
chaotic behavior may be the universal feature in the seemingly
random light curves found in many sources. It is expected to find a
common nonlinear dynamical origin to explain this chaotic behavior.
This dynamical origin may be the nonlinear temporal evolution of the
magneto-hydrodynamic flow of the inner accretion disk (Misra et
al.~2004, 2006). It also may come from the turbulent motion in the
gaseous cloud around the object (Li \& Xiao 2000).

Some authors (Uttley et al.~2005) also consider that the dynamical
chaos is not required to explain the data. But our results together
with Misra R. et al. cogently prove that the seemingly random light
curves found in some sources are indeed chaotic.  We expect that our
results are confirmed further by using other chaotic and nonlinear
time-series analysis.

\begin{acknowledgements}

I would like to thank Professor Meng Ta-chung for his contribution
to my knowledge of nonlinear dynamics. I would also like to thank
Wang Xiao-Dong and Yu Yun-Wei for their critical readings of the
manuscript. At last,I want to give my especial thanks to Professor
Wang En-Ke for his enthusiastic help and encouragement.

\end{acknowledgements}

\label{lastpage}

\end{document}